\documentclass[preprint,aps,prc,superscriptaddress,showpacs,floatfix]{revtex4}
\usepackage{graphicx}

\begin{document}

\title{Effect of resonance decays on hadron elliptic flows}

\author{V. Greco}
\affiliation{Cyclotron Institute and Physics Department, Texas A\&M 
University, College Station, Texas 77843-3366}
\author{C. M. Ko}
\affiliation{Cyclotron Institute and Physics Department, Texas A\&M 
University, College Station, Texas 77843-3366}

\date{\today}

\begin{abstract}
Within the quark coalescence model, we study effects of resonance decays,
and of the quark momentum distribution in hadrons, on the elliptic 
flows of stable hadrons.  We find that, with the exception 
of rho-meson decays, the resonance decays could have a significant effect
on pion elliptic flow. However, most secondary pions stem from the 
rho-meson decays, resulting in a reduced effect of resonance decays 
on their flow. Proton and kaon flows as well as the lambda flow are, 
however, not much affected by resonance decays. The distribution of 
quark momentum in hadrons also influences their elliptic flows,
leading to a better agreement with experimental flow data when 
compared with the naive quark coalescence model, which only allows 
quarks with equal momentum to form a hadron.
\end{abstract}
\pacs{25.75.-q,25.75.Ld,25.75.Nq}

\maketitle

\section{introduction}

A useful observable for understanding both the dynamics of heavy-ion 
collisions and the properties of produced hot and dense matter is the 
elliptic flow of hadrons, i.e., their momentum anisotropy in the 
transverse plane perpendicular to the beam directions.  For heavy-ion 
collisions at the Relativistic Heavy Ion Collider (RHIC), it has been 
shown in studies based on hydrodynamical models \cite{tean,kolb,huov} 
that hadron elliptic flows are sensitive to the equation of state 
of the quark-gluon plasma produced during the initial stage of 
the collision. In transport models \cite{zhang,moln1,lin1}, hadron 
elliptic flows are also shown to depend on the parton scattering cross 
sections in initial partonic matter. For hadrons with high transverse 
momentum, their elliptic flow can further provide information on 
the energy density of initial hot matter \cite{gyulv2}. Experimentally, 
hadron elliptic flows in Au+Au collisions at $\sqrt{s_{NN}}= 130$ and $200$ 
GeV have been studied as functions of pseudo-rapidity \cite{man}, 
centrality \cite{ack,esu}, and transverse momentum \cite{esu,adl}. 
Furthermore, the elliptic flows of identified particles have 
been measured, and except for pions they essentially follow the quark 
number scaling, i.e., the dependence of hadron elliptic flows
on hadron transverse momentum becomes similar if both are divided 
by the number of constituent quarks in a hadron, i.e., two for mesons 
and three for baryons. The scaling of hadron elliptic flows according 
to their constituent quark numbers has a simple explanation in the 
naive quark coalescence model \cite{volo}, in which the meson elliptic flow at 
certain transverse momentum is given by twice the quark 
elliptic flow at half the meson transverse momentum, while that of 
baryons is given by three times the quark elliptic flow at one third 
of baryon transverse momentum. In more realistic quark coalescence models 
\cite{greco2,fries2,hwa,greco,fries} that take into account the 
momentum distribution of quarks in hadrons, the scaled hadronic 
elliptic flows are, however, expected to be smaller than that of 
partons \cite{lin2}, and this may lead to a violation of the 
quark number scaling of hadron elliptic flows.

In this paper, we study how the elliptic flows of pions and other 
stable hadrons are affected by decays of resonances, such as 
$\rho \rightarrow 2\,\pi$, $\omega \rightarrow 3\,\pi$, 
$K^* \rightarrow K\,\pi$, and $\Delta \rightarrow N\,\pi$. For most
hadrons, such as the proton, kaon, and lambda, including 
contributions from resonance decays do not affect much their elliptic 
flows. This is, however, different for pions as the elliptic flow
of pions from the decays of most resonances, except the $\rho$ meson, 
are significantly different from that of directly produced pions. 
This effect is found to account for half of the observed deviation 
of pion elliptic flow from the quark number scaling. Taking into 
account the quark momentum distribution in hadrons, which makes it 
possible for quarks with different momenta to coalescence into
hadrons, we find that the other half of the observed deviation can 
also be largely explained.

The paper is organized as follows. In Sect. \ref{scaling}, we
review observed approximate quark number scaling of the elliptic
flows of identified hadrons and interpret it with the naive 
quark coalescence model. The effect of resonance decays on the 
elliptic flows of stable hadrons is then studied in Sect. 
\ref{resonances}. In Sect. \ref{momentum}, we further investigate
the effect on hadron elliptic flows due to quark momentum distribution
inside hadrons. Finally, a summary is given in Sect.\ref{summary}.

\section{scaling of hadron elliptic flows}\label{scaling}

\vspace{1cm}

\begin{figure}[th]
\includegraphics[height=3in,width=4.0in,angle=0]{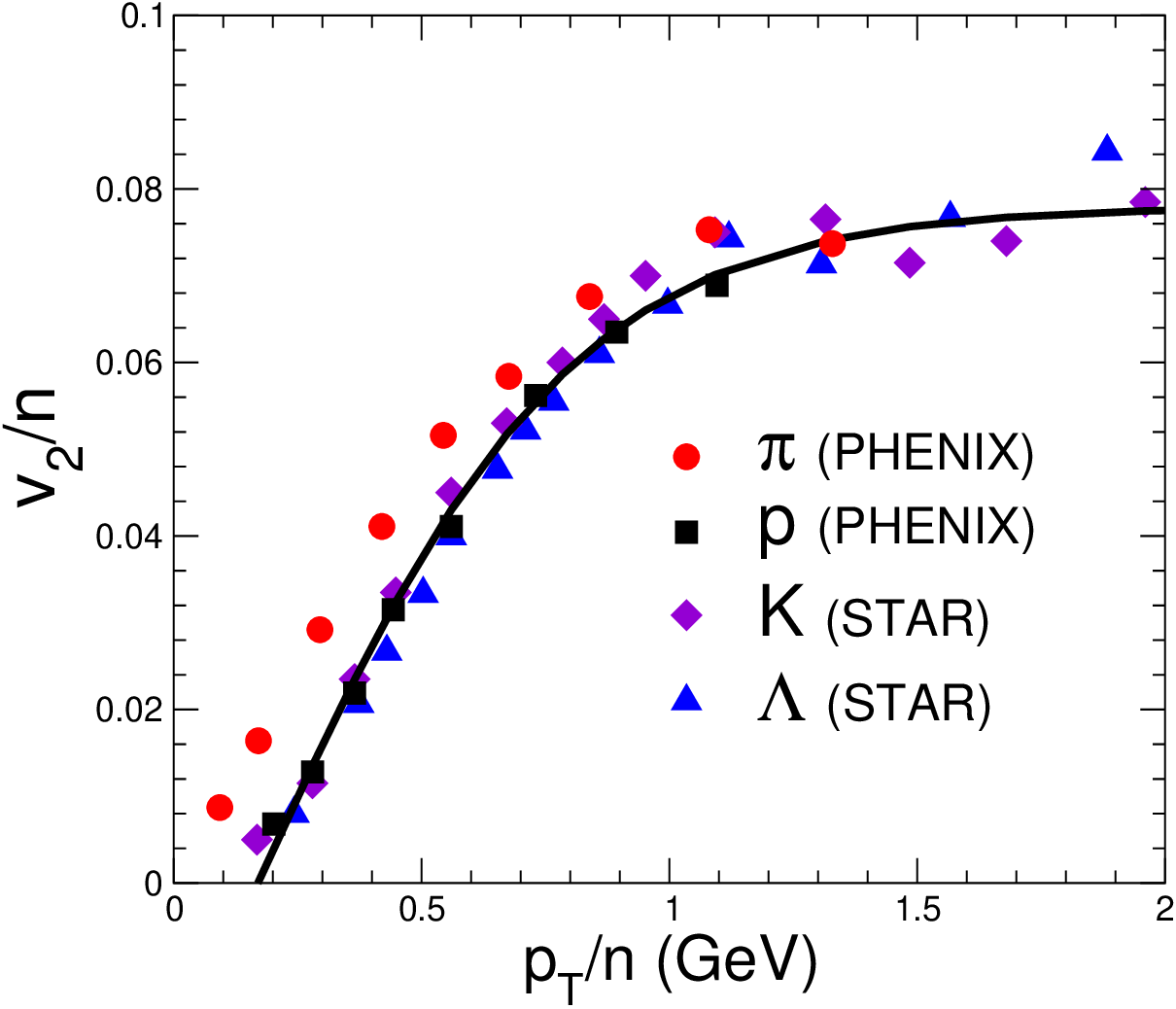}
\caption{(Color online)
Experimental scaled elliptic flows as functions of scaled momentum for 
pions (circles) and protons (squares) from the PHENIX Collaboration
\cite{sadl} as well as kaons (diamonds) and lambdas (triangles) 
from the STAR Collaboration \cite{sor} in Au+Au collisions 
at $\sqrt{s_{NN}}=200$ GeV. The solid line given by Eq.(\ref{v2q}) 
is a fit to the experimental data for hadrons other than pions.} 
\label{fig1}
\end{figure}

In Fig.\ref{fig1}, we first show the experimental data from Au+Au 
collisions at $\sqrt{s_{NN}}=200$ GeV for the elliptic flows of 
pions (circles) and protons (squares) from the PHENIX Collaboration
\cite{sadl} as well as $K^0_s$ (diamonds) and lambdas (triangles) 
from the STAR Collaboration \cite {sor} scaled by their constituent 
quark numbers as functions of their transverse momentum per 
quark \cite{comment}. Except for pions, the scaled elliptic flows 
of other stable hadrons are indeed similar and can be described by the 
naive quark coalescence model \cite{volo} using the quark elliptic flow 
\begin{equation}
v_{2,q}(p_T)=v_0\,{\rm tanh}(\alpha\, p_T +\phi),
\label{v2q}
\end{equation} 
with $v_0=0.078$, $\alpha=1.59$ and $\phi=-0.27$. In this model, the
elliptic flows of mesons and baryons are related to that of quarks by 
$v_{\rm 2,M}(p_T)\approx 2\,v_{2,q}(p_T/2)$ and 
$v_{\rm 2,B}\approx 3\,v_{2,q}(p_T/3)$, respectively, where small 
higher-order corrections are neglected \cite{kolb1}. The resulting 
scaled hadron elliptic flows are shown in Fig.\ref{fig1} by the 
solid line, which is seen to give a good description of the
experimental data for stable hadrons other than pions.

\section{effect of resonance decays}\label{resonances} 

A large fraction of stable hadrons such as pions, kaons, lambdas, 
and protons, detected
in heavy ion collisions are from resonance decays. Although, 
in the coalescence model, the elliptic flow of both direct hadrons 
and resonances essentially follows the quark number scaling, hadrons  
from resonance decays may have elliptic flow that deviate from 
this scaling.  To see the effect of resonance decays on the elliptic 
flow of stable hadrons, we use the coalescence model of
Ref.\cite{greco2} to generate their transverse momentum distributions. 
Specifically, soft partons with transverse momentum below 2 GeV/$c$, 
which are taken to have constituent quark masses, are assumed to have 
thermal spectra at temperature $T=170$ MeV with a collective radial
flow increasing linearly with radial distance up to a maximum flow 
velocity of 0.5$c$, while hard partons above 2 GeV/$c$ are given by 
the quenched minijet partons with power-like spectra. Furthermore, 
the volume of partonic matter at hadronization is taken to be $V=900$ 
fm$^{3}$, leading to a parton density of about $1.15$ fm$^{-3}$. The 
momentum spectra of hadrons are then obtained from the overlap between
the parton distribution functions with the Wigner distribution
functions of hadrons, which we evaluate using the Monte Carlo method
of Ref.\cite{greco2}. The hadron Wigner distribution functions are 
taken to have a momentum width of $\Delta_p=0.24$ GeV/$c$ for mesons 
and 0.35 GeV/$c$ for baryons. These parameters have been shown in 
Ref.\cite{greco2} to describe very well the measured transverse momentum
spectra of pions, kaons, and protons. 

\subsection{Transverse momentum spectra}

\begin{figure}[th]
\includegraphics[height=5in,width=4.0in]{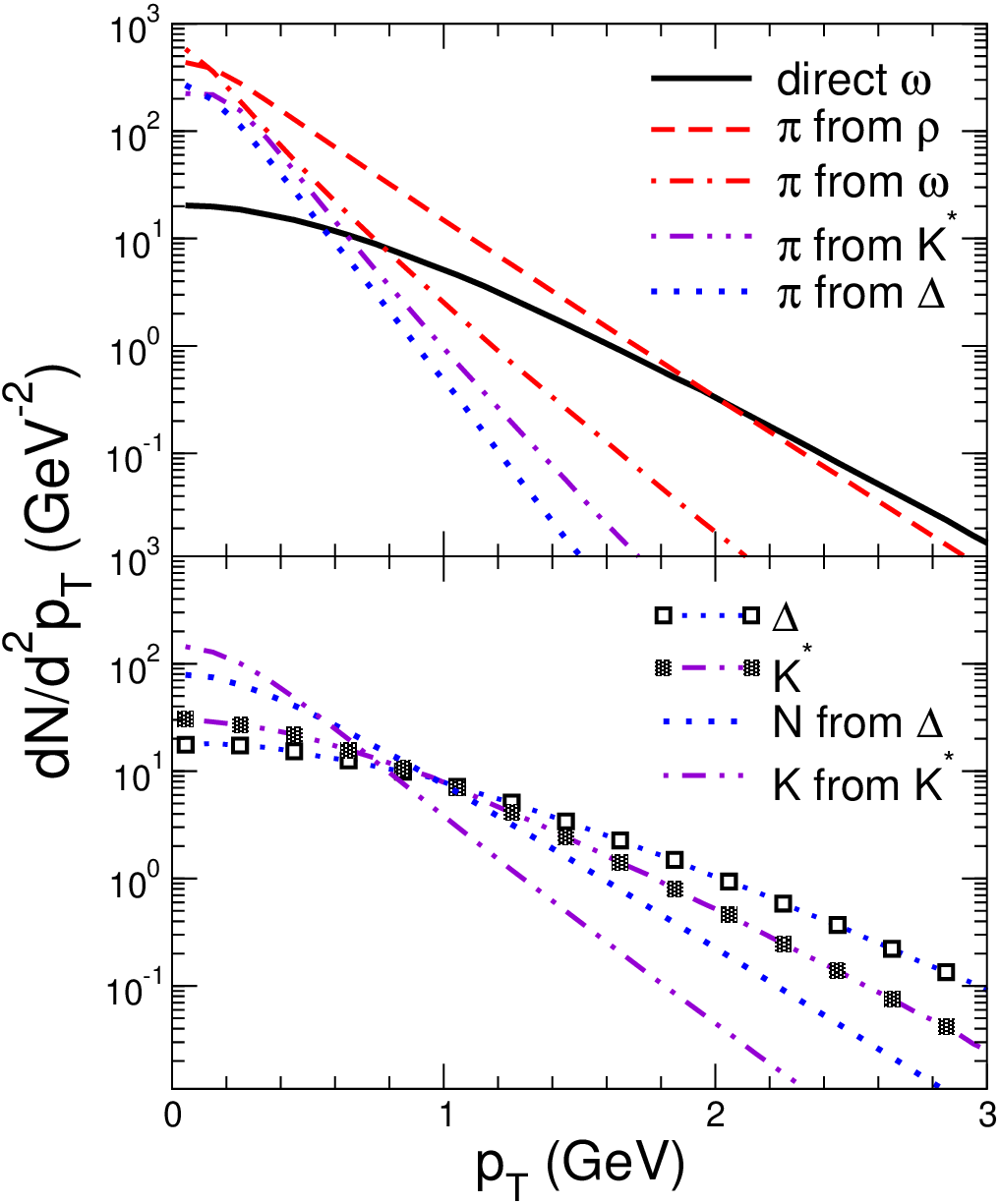}
\caption{(Color online)
Transverse momentum spectra of hadrons. Upper panel:
spectra of pions from decays of $\rho$ mesons (dashed line), 
$\omega$ mesons (dash-dotted line), $K^*$ (dash-dot-dotted line), 
and $\Delta$ (dotted line). The spectrum of $\omega$ mesons is shown 
by the solid line. Lower panel: spectra of $\Delta$ (dotted line with 
open squares) and $K^*$ (dash-dot-dotted line with 
shaded squares) together with those of nucleons (dotted line) and $K$
(dash-dot-dotted line) from their decays.}
\label{fig2}
\end{figure}

We first show in Fig.\ref{fig2} the transverse momentum spectra of 
various hadrons obtained from the quark coalescence model of Refs.
\cite{greco,greco2}. In the upper panel, the spectrum of $\omega$ mesons 
is given by the solid line, and that of pions from $\omega$ decays is 
given by the dash-dotted line and is seen to be steeper than the original 
$\omega$ meson spectrum. The same is true for the transverse momentum 
spectra of pions from decays of $\rho$ meson (dashed line), 
$K^*$ (dash-dot-dotted line), and $\Delta$ (dotted line), i.e., they
are steeper than the transverse momentum spectra of original
resonances.  These pions, except from $\rho$ meson decays, also have 
steeper transverse momentum spectra than that of pions from $\omega$ 
decays. We note that in the coalescence model more pions are
produced from $\rho$ meson decays than from the decays of other 
resonances and from direct recombination of quarks and antiquarks.

In the lower panel of Fig.\ref{fig2}, we show the initial transverse
momentum spectra of $\Delta$ (dotted line with open squares) and $K^*$ 
(dash-dot-dotted line with shaded squares) as well as those of nucleons
(dotted line) and $K$ (dash-dot-dotted line) from their decays.  The 
differences between the slope parameters of nucleon and $K$ transverse 
momentum spectra and those of the parent $\Delta$ and $K^*$ are smaller
than those between the slope parameters of the transverse momentum
spectra of pions and their parent resonances. Our results on the 
transverse momentum spectra of resonances and their decay products 
are consistent with those using the thermal model \cite{soll}, 
although there are differences in detail due to different hadron 
abundances and the inclusion of flow effects in our study.

\subsection{Elliptic flows}

\begin{figure}[th]
\includegraphics[height=5in,width=4.0in]{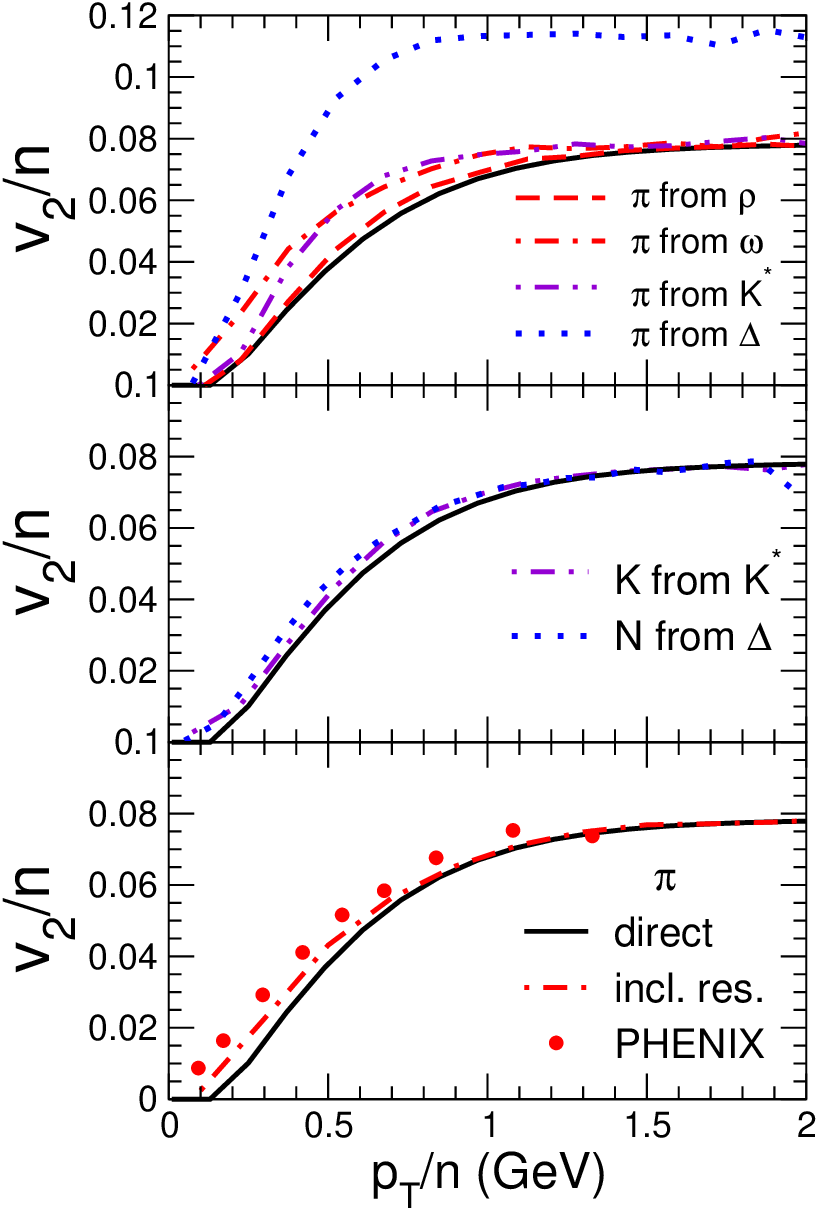}
\caption{(Color online)
Scaled hadron elliptic flows as functions of scaled momentum.  Upper panel: 
pions from decays of $\rho$ mesons (dashed line), $\omega$ 
(dash-dotted line), $K^*$ (dash-dot-dotted line), and $\Delta$ (dotted
line). Middle panel: $K$ from $K^*$ decays (dash-dot-dotted line) 
and protons from $\Delta$ decays (dotted line). Lower panel: pions 
including resonance decays (dash-dotted line) and from experimental 
data (solid circles).  The solid line in all panels represents the scaled 
elliptic flow (Eq.(\ref{v2q})) obtained from fitting the experimental 
data for stable hadrons other than pions.}
\label{fig3}
\end{figure}

To study the effect of resonance decays on the elliptic flow of stable
hadrons, we assume that all resonances produced via coalescence 
have the scaled elliptic flow given by Eq.(\ref{v2q}), as in 
the naive quark coalescence model. Since a decay particle at given 
transverse momentum arises from a resonance at higher momentum with 
identical elliptic flow,  decay products are expected to have a larger
elliptic flow than directly produced particles.
For low-momentum particles, however, the decay process, which is 
isotropic in the rest frame 
of the resonance, reduces the momentum anisotropy (and thus the 
elliptic flow). At very low momentum, this 
effect can be large enough to render the elliptic flow of particles 
from resonance decays even smaller than that of directly produced 
ones, as first pointed out in Ref.\cite{hira}.  On the other hand,
particles at large momentum are essentially aligned with the parent 
resonances, and consequently tend to preserve the 
elliptic flow of the parent resonances. 
These features are indeed seen in Fig.\ref{fig3}, where the 
scaled elliptic flow of stable hadrons from resonance decays 
is shown. 

In the upper panel of Fig.\ref{fig3}, we show the scaled elliptic flows of 
pions from decays of various resonances. For pions from $\rho$ meson 
decay, the scaled elliptic flow, given by the dashed line, differs 
very little from that of directly produced pions, which in the naive 
quark coalescence model follows the quark number scaling given by 
the solid line. The slight enhancement of the pion elliptic flow 
at larger transverse momenta due to rho meson decays has also been 
seen previously in studies based on the thermal model \cite{bronio}.
The dash-dotted line is the scaled pion elliptic flow from the decay 
of $\omega$ mesons. It has a larger value than that of directly 
produced pions as the average transverse momentum of these pions is 
even smaller than that of pions from $\rho$ meson decays as shown 
in Fig.\ref{fig2}. Similarly, the elliptic flow of pions from $K^*$ 
decays (dash-dot-dotted line) is large and comparable 
to that of pions from $\omega$ decays. For pions from $\Delta$ decays 
(dotted line), their elliptic flow is even larger than that for pions 
from $\omega$ and $K^*$ decays. Therefore, pions from resonance decays 
can lead to a deviation of the final pion elliptic flow from the 
quark number scaling.  As shown in the lower panel, the elliptic 
flow of pions including those from resonance decays (dash-dotted line) 
is indeed closer to the experimental data (solid circles) than that of 
directly produced pions (solid line) expected from the naive quark 
coalescence model. The effect of resonance decays on the pion elliptic
flow is thus not small and can roughly account for half of the observed 
deviation of pion elliptic flow from the quark number scaling.

In the middle panel of Fig.\ref{fig3}, we show the elliptic flows of
$K$ (dash-dot-dotted line) and nucleons (dotted line) from decays of 
$K^*$ and $\Delta$, and they are seen to differ only slightly 
from those of directly produced ones, which again follow more or less 
the quark number scaling given by the solid line. Resonance decays 
thus do not destroy the observed approximate scaling of proton,
kaon, and lambda elliptic flows according to their constituent quark
number. The difference between the elliptic flows of hadrons and their 
decay products has also been studied for charmed mesons \cite{greco1}. 
It was found that the elliptic flow of electrons from charmed meson
decays was also slightly larger than that of charmed mesons. 

The above results are obtained without taking into account the
width of resonances. Because of their finite widths, resonances are 
produced with masses given by the Breit-Wigner distribution. Since the
momenta of pions from decays of resonances with masses below (above)
their peak mass are smaller (larger) than the momenta of pions from 
decays of resonances with peak mass, the elliptic flows of these pions
are larger (smaller). As a result, the increased elliptic flow of 
pions from resonances with masses below their peak value
is canceled by the decreased elliptic flow of pions from resonances
with masses above their peak values. Including the effect of
resonance width is thus not expected to affect the above results.
We have checked this for the $\rho$ meson, which dominates pion production
in the coalescence model and has a width of 150 MeV, and indeed we have found
that the resulting pion elliptic flow is essentially the same as the one
without including the rho meson width. This remains so even if we 
increase the $\rho$ meson width to 300 MeV due to possible collision 
broadening in dense medium \cite{rapp}.

Since a lower rho meson mass leads to a larger pion elliptic flow, 
it is interesting to note that the remaining deviation of scaled pion 
elliptic flow from the quark number scaling could be explained if the 
rho meson mass is reduced from 770 MeV/$c^2$ to 500 MeV/$c^2$. A dropping rho
meson mass in dense medium has been suggested to relate to the partial 
restoration of chiral symmetry \cite{brown}. It has also been shown to give a 
possible explanation \cite{gqli,cassing} for the enhanced production 
of low mass dileptons observed in heavy ion collisions at the SPS 
\cite{agakichiev}. More recently, the $\rho$ meson mass extracted from 
the two-pion invariant mass spectrum in heavy ion collisions at 
RHIC was found about 70 MeV/$c^2$ less than its value in free space 
\cite{fachini}. However, before one can draw such an conclusion, 
it is important to investigate other conventional effects 
that have been neglected in the naive quark coalescence model, and
this is discussed in the next section.

\section{effect of quark momentum distribution in hadrons}\label{momentum}

\begin{figure}[th]
\includegraphics[height=3in,width=4.0in]{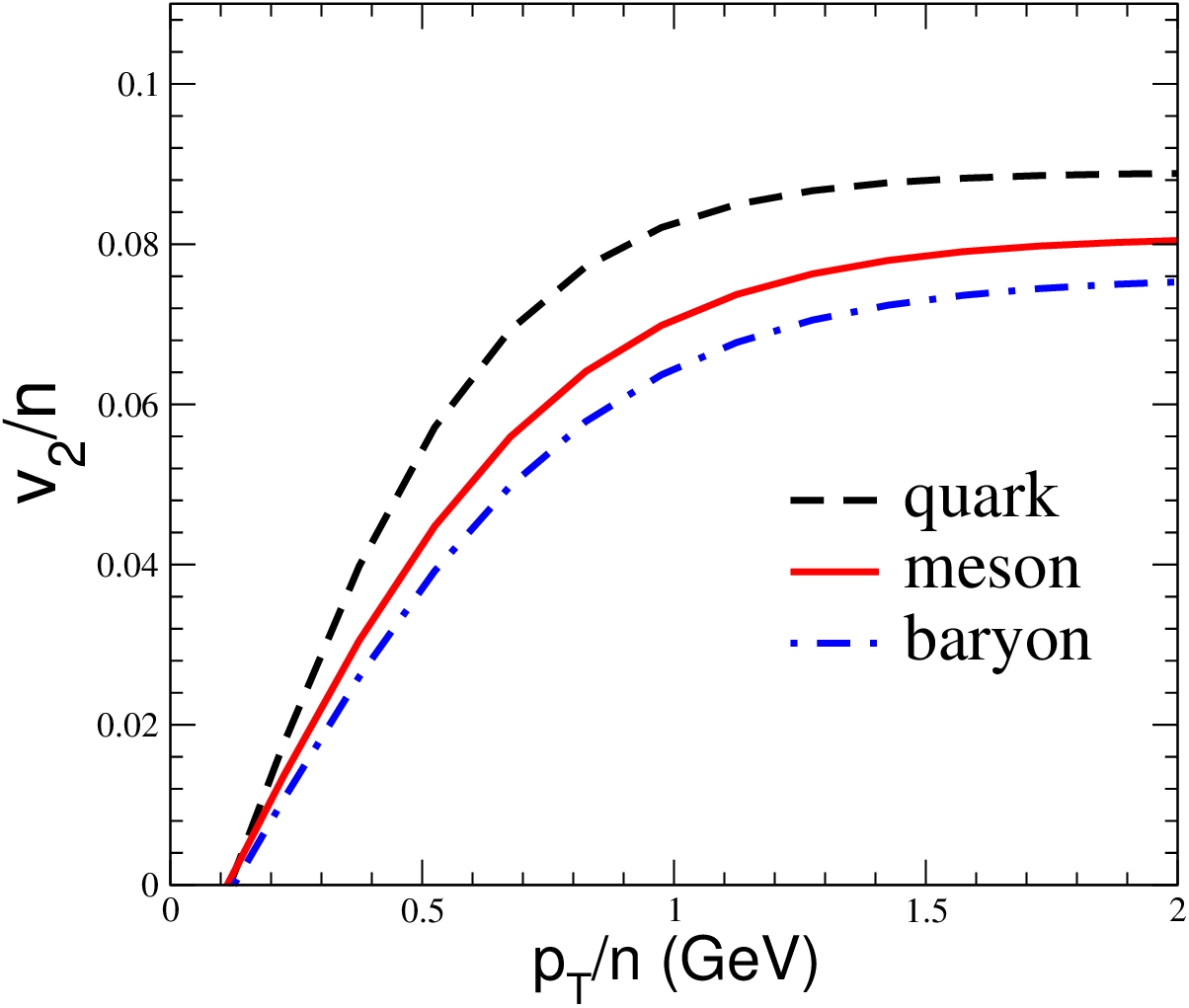}
\caption{(Color online)
Scaled elliptic flows of quarks (solid line), meson (dashed line)
and baryons (dotted line) in a realistic quark coalescence model.}
\label{fig4}
\end{figure}

The scaling of hadron elliptic flows with their constituent
quark number only holds in the naive quark coalescence model that 
requires mesons at transverse momentum $p_T$ to be formed 
from partons at momentum $p_T/2$ and baryons from quarks at 
momentum $p_T/3$. In more realistic coalescence models, hadron 
production is expressed in terms of the overlap of parton
distributions with the Wigner functions of formed hadrons, so
quarks with different momenta can coalescence into hadrons \cite{greco2}.
As shown in Ref.\cite{lin2}, in which only collinear quarks can form 
hadrons, the effect due to the momentum distribution of quarks in
hadrons can lead to a breaking of quark number scaling if the quark elliptic 
flow does not depend linearly on the transverse momentum. To study
this effect quantitatively, we again use the Monte Carlo method of 
Ref.\cite{greco2} to evaluate the coalescence integral so that 
non-collinear quarks can also form hadrons. To obtain scaled hadron 
elliptic flows that are comparable to the measured ones, we further 
take quarks to have an elliptic flow similar to Eq.(\ref{v2q}) but 
with $v_0=0.089$, $\alpha=1.89$, and $\phi=-0.22$. This gives a larger 
quark elliptic flow, shown by the solid line in Fig.\ref{fig4}, 
than the quark elliptic flow in Fig.\ref{fig1}. The resulting scaled 
meson (dashed line) and baryon (dotted line) elliptic flows are shown 
in Fig. \ref{fig4}, and they are smaller than the quark elliptic
flow. This effect is larger for baryons than for mesons, leading 
thus to a scaled elliptic flow that is smaller for baryons than for mesons. 


\begin{figure}[th]
\includegraphics[height=3in,width=4.0in]{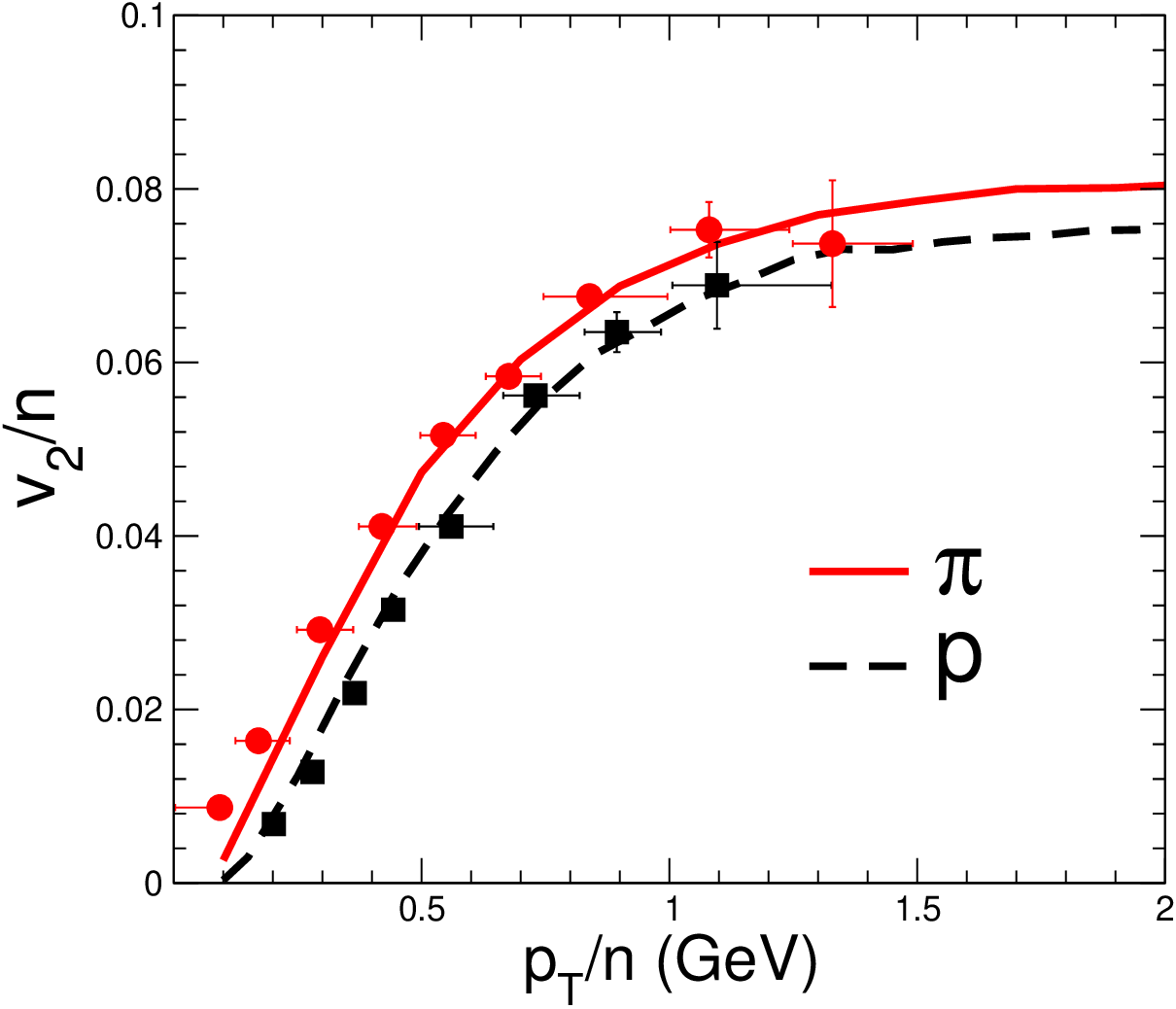}
\caption{(Color online)
Scaled pion (dashed line) and proton (dotted line) elliptic flows in 
the coalescence model that includes effects due to both
resonance decays and the quark momentum distribution in hadrons.
Experimental data are shown by circles for pions and squares for 
protons \cite{sadl}.}
\label{fig5}
\end{figure}

Including also the contribution from decays of resonances, the resulting 
scaled pion (dashed line) and proton (dotted line) elliptic flows are 
shown in Fig.\ref{fig5} together with the experimental data \cite{sadl}
for pions (circles) and protons (squares). The observed difference in 
the scaled pion and proton elliptic flows can be largely
explained as already shown in Ref.\cite{greco2}. It was also shown
in Ref.\cite{greco2} that the more realistic coalescence model could 
describe the measured elliptic flow of kaons and lambdas if the 
underlying elliptic flow of strange quarks is taken to be slightly 
different from that of light quarks. Therefore, a combination of the effects 
due to the quark momentum distribution in hadrons and resonance decays 
in the coalescence model can lead to the observed quark number scaling of 
hadron elliptic flows as well as the violation of the pion elliptic 
flow from this scaling behavior.

The measured hadron elliptic flows including that of pions for 
transverse momentum $p_T \leq 1.5$ A GeV have also been explained by  
the hydrodynamical model \cite{huov}, which predicts that hadron
elliptic flows are affected by their masses with heavier ones having
larger elliptic flows. As a result, the observed violation of 
scaled pion elliptic flow from the quark number scaling can be attributed
to its small mass. Since the phi meson mass is similar to that of
proton but is a meson, its elliptic flow is similar to that of proton 
in the hydrodynamical model and is thus expected to deviate also from 
the quark number scaling. This is in contrast to the prediction of the quark 
coalescence model that the phi meson elliptic flow should satisfy the quark
number scaling as well \cite{greco2}. Measurement of the phi meson elliptic 
flow in relativistic heavy ion collisions is thus important for 
testing the different predictions from these two models \cite{nona1}.

In the above study, we have neglected the effect of rescattering
on hadron elliptic flows. This is justified as studies based on
transport models have shown that elliptic flows in heavy ion collisions
are generated during the initial stage of collisions when both spatial 
and pressure anisotropies are appreciable, and their values 
remain essentially unchanged during hadronic evolution \cite{lin1,chen}.
Only for resonances with transverse momentum $p_T/n\simeq 1.5$ GeV, 
where $n$ is the number of constituent quarks in a resonance, 
their elliptic flows may be affected by hadronic rescattering effect
\cite{nona}. 

\section{summary}\label{summary}

In summary, we have studied the effect of resonance decays on 
the elliptic flows of stable hadrons, particularly that of pions.
For heavier hadrons such as $K$ and protons, their elliptic flows  
are not much affected by decays of $K^*$ and $\Delta$. This is
different for pions as the elliptic flows of pions from resonance
decays, except the $\rho$ meson, are larger than that of directly 
produced pions. Since $\rho$ meson decays are more important 
than other resonances in contributing to the final yield of pions, 
the final pion elliptic flow is, however, only moderately affected by 
resonance decays, leading to a deviation of the pion elliptic flow 
from the quark number scaling that is about half the observed one.  
Including the effect of quark momentum distribution in hadrons not 
only reduces the hadron elliptic flows but also leads to additional 
deviation of the scaled pion elliptic flow from that of protons. 
The quark coalescence model that takes into account both effects 
of resonance decays and quark momentum distribution in hadrons 
thus can give a good description of not only the hadron transverse 
momentum spectra but also their elliptic flows. The agreement between 
theory and experiments at low momenta may, however, be fortuitous 
as the coalescence model, which ignores energy conservation, 
is likely to be too crude for hadrons with low momenta. To improve 
the model, one needs to take into account the effect of other 
partons in the system in order to balance the energy mismatch 
during quark coalescence, which is beyond the scope of present study. 

\section{acknowledgments}

We are grateful to Lie-Wen Chen, Ralf Rapp, and Nu Xu for helpful discussions.
This paper was based on work supported in part by the US National
Science Foundation under Grant No. PHY-0098805 and the Welch
Foundation under Grant No. A-1358. V.G. was also supported by 
the National Institute of Nuclear Physics (INFN) in Italy.
 
\bigskip

\noindent{{\it Note added:} After submission of this paper, we were
aware of a paper by X. Dong, S. Esumi, P. Sorensen, and N. Xu
\cite{dong} which also reaches a similar conclusion that 
resonance decays affect appreciably the pion elliptic flow.


\begin{thebibliography}{99}
\bibitem{comment}Here and elsewhere in the paper, we use $\pi$ for
both $\pi^+$ and $\pi^-$, $p$ for both $p$ and $\bar p$. The same
is for other particles, i.e., they denote both particles and 
their antiparticles.
\end{thebibliography}
\end{document}